\documentclass[english,aps,prl,superscriptaddress,twocolumn]{revtex4-1}
\usepackage[T1]{fontenc}
\usepackage[latin9]{inputenc}
\usepackage{xcolor}
\usepackage{amsmath}
\usepackage{amssymb}
\usepackage{graphicx}
\usepackage{esint}
\PassOptionsToPackage{normalem}{ulem}
\usepackage{ulem}

\makeatletter

\providecolor{lyxadded}{rgb}{0,0,1}
\providecolor{lyxdeleted}{rgb}{1,0,0}
\DeclareRobustCommand{\lyxsout}[1]{\ifx\\#1\else\sout{#1}\fi}

\usepackage{slashed}

\makeatother

\usepackage{babel}
\begin{document}

\title{Torsional Responses and Liouville Anomaly in Weyl Semimetals with Dislocations}

\author{Ze-Min Huang}

\affiliation{School of Physics, Sun Yat-Sen University, Guangzhou 510275, China}

\affiliation{Department of Physics, University of Illinois, 1110 W. Green St. Urbana,
IL 61801 USA }

\author{Longyue Li}

\affiliation{School of Physics, Sun Yat-Sen University, Guangzhou 510275, China}

\author{Jianhui Zhou}
\email{jhzhou@hmfl.ac.cn}

\selectlanguage{english}%

\affiliation{Anhui Province Key Laboratory of Condensed Matter Physics at Extreme
Conditions, High Magnetic Field Laboratory, Chinese Academy of Sciences,
Hefei 230031, Anhui, China}

\author{Hong-Hao Zhang}
\email{zhh98@mail.sysu.edu.cn}

\selectlanguage{english}%

\affiliation{School of Physics, Sun Yat-Sen University, Guangzhou 510275, China}

%\date{\today}
\begin{abstract}
Weyl nodes in three-dimensional Weyl semimetals break the Liouville
equation, leading to the Liouville anomaly.
Here we present a new approach to derive the semiclassical action and equations of motion
for Weyl fermions in the presence of electromagnetic fields and torsions
from the quantum field theory: combining the Wigner transformation
with band projection operation.
It has been shown that the Liouville anomaly, including a new pure torsion anomaly term, a mixing term between the electromagnetic fields and torsions as well as the conventional chiral anomaly, entirely differs from the counterpart of axial gauge fields.
We find various torsional responses and reproduce the chiral vortical effect and the torsional chiral magnetic effect.
A new torsion modified anomalous Hall effect due to the mixing term in the Liouville anomaly is predicted and its implementation is also discussed.
Therefore, our work not only provides new insights into the torsional responses for Weyl fermions but also acts as a starting point to investigate their topological responses.
\end{abstract}
%%%%%%%%%%%%%%%%%%%%%%%%%%%%%%%%%%%%%
\maketitle
\section{Introduction}
Quantum anomalies, the breaking of classical
symmetries by quantum fluctuations, have attracted much attention
in condensed matter physics due to the deep connection with topology
\citep{fujikawa2004oxford,bertlmann2000oxford}. Many exotic responses
of topological phases of matter can be understood in the language
of quantum anomalies, including topological insulators \citep{witten2016rmp,ryu2012prb}
and topological semimetals \citep{zyuzin2012prb,WangZ2013PRB,liu2013PRB}. Recently, the chiral
anomaly in the context of three dimensional Weyl and Dirac semimetals
has led to rich physical phenomena \citep{hosur2013crp,weng2016jpcm,armitage2018rmp,burkov2018arcp},
such as the chiral magnetic effect (CME) \citep{vilenkin1980prd,fukushima2008prd,Grushin2012PRD,son2012prl,zyuzin2012prb,zhou2013cpl,Goswami2013PRB,vazifeh2013prl,landsteiner2014PRB,chang2015prb2},
the negative longitudinal magnetoresistance \citep{son2013prb}, the
nonlocal transport \citep{Parameswaran2014PRX}, the giant planar
Hall effect \citep{Burkov2017PRBgphe,Nandy2017PRL}, and the unconventional
collective excitations \citep{ZhouJH2015PRB,GorbarPRL2017,Song2016PRB,Rinkel2017PRL,Liu2017PRL},
some of which have been observed experimentally \citep{Kim2013PRL,Huang2015PRX,xiong2015science,li2015naturecom,li2016naturecom,li2016naturephy,MWu2017phe}.

Historically, the chiral anomaly was first derived by use of the perturbation
method \citep{adler1969pr,bell1969nca}, later by the Fujikawa's path
integral method \citep{fujikawa2004oxford} and from the transport
of the chiral zeroth Landau level \citep{nielsen1983plb}. Recently,
the Berry curvature modified semiclassical equations of motion are
used to derive the equation of the chiral anomaly \citep{stephanov2012prl}.
Within the framework of the semiclassical equations of motion \citep{xiao2010rmp},
the chiral anomaly equation manifests itself as breaking the conservation
of the phase-space current or the Liouville equation, which is also
dubbed as the Liouville anomaly \citep{stephanov2012prl,stone2013prd,dwivedi2013iop}.
Compared with Fujikawa\textquoteright s method, the absence of ultraviolet
cut-off in the Liouville anomaly can be traced back to the charge
pumping between Weyl nodes with opposite chirality. In the presence
of dislocations or temperature gradients, gravity with torsion would
emerge. A non-vanishing torsion contributes the Nieh-Yan term to the chiral
anomaly equation~\citep{nieh1982jmp}, leading to novel geometrical responses for Weyl fermions~\citep{you2016prb,chernodub2017prb}.

In fact, the construction of the semiclassical equations of motion with
torsions is highly nontrivial. Both the wave-packet approach \citep{sundaram1999prb}
and the chiral kinetic theory \citep{stephanov2012prl} are based
on the Hamiltonian mechanics, while the Hamiltonian from the curved-spacetime
Dirac equation is tortured by the Hermiticity problem \citep{parker1980prd}.
Although a Hermitian Hamiltonian can be obtained from some careful
manipulations \citep{huang2009prd}, it would be too cumbersome
for our purposes. In addition, semiclassical chiral kinetic theory
can also be derived from quantum field theory, but it only keeps valid
in the homogeneous limit \citep{son2013prd}. Therefore, a new method
for deriving the semiclassical equations of motion with torsions is
highly desirable and crucial to investigations of the related topological responses.

In this paper, we develop a new formalism to derive the semiclassical
action and equations of motion for Weyl fermions in the presence of
electromagnetic fields and torsions from the quantum field theory:
combining the Wigner transformation with band projection operation.
The relevant Liouville anomaly consists of a new pure torsion anomaly term, and a term mixing the electromagnetic fields and torsions, in addition to the conventional chiral anomaly.
Various novel responses are obtained, such as the chiral vortical effect and the newly proposed torsional chiral magnetic effect.
Meanwhile, we find a new torsion modified anomalous Hall effect (AHE) from the mixed term in the Liouville anomaly and discuss its implementation in Weyl semimetals with broken time reversal symmetry such as Co$_{3}$Sn$_{2}$S$_{2}$. In addition, we find that the chiral zero modes localized in dislocations can be cancelled by the compensation of those states from the bulk via the Callan-Harvey Mechanism.

The rest of this paper is organized as follows.
In Sec. \ref{model}, the Lagrangian density for Weyl fermions in the presence of torsions is introduced.
In Sec. \ref{Green}, we derive the one-band effective Green's function by combining the band projection with Wigner's transformation.
In Sec. \ref{action and EOM}, the semiclassical dynamics in the presence of both torsions and electromagnetic fields are derived.
In Sec. \ref{responses_anomaly}, we consider torsional responses and derive the anomaly equations within the semiclassical formalism. We also predict a torsion modified AHE and discuss its experimental implementation.
In sec. \ref{conclusion}, the main results of this paper are summarized. Finally, we give the detailed calculations in the Appendices.

%%%%%%%%%%%%%%%%%%%%%%%%%%%%%%%%%%%%%
\section{Model}\label{model}
In the presence of dislocations, the corresponding deformation of media is described
by the displacement vectors $u^{a}\left(x\right)$, where the superscript
$a=0,\thinspace1,\thinspace2,\thinspace3$ denotes locally flat spacetime
coordinates with the metric tensor $\eta_{ab}=\text{diag}\left(1,\thinspace-1,\thinspace-1,\thinspace-1\right)$~\cite{weakdisp}.
That is, under a lattice deformation, lattice coordinates are shifted, i.e. $x\rightarrow x+u$. So there is $\partial_{\mu}=e^{a}_{\mu}\partial_{a}$, where $e_{\mu}^{a}$
is the vielbein, i.e. $e_{\mu}^{a}=\delta_{\mu}^{a}+\partial_{\mu}u^{a}$, and $\mu=0,\thinspace1,\thinspace2,\thinspace3$ denotes
the curved spacetime coordinates (or lab coordinates) with the metric
tensor $g_{\mu\nu}=e_{\mu}^{a}\eta_{ab}e_{\nu}^{b}$. The action has the form \citep{taylor2011prl,taylor2013prd}
\begin{equation}
S=\frac{1}{2}\int d^{4}x\left|\det e_{\mu}^{a}\right|\left[\bar{\Psi}e_{a}^{\mu}\gamma^{a}\left(i\partial_{\mu}\Psi\right)-\left(i\partial_{\mu}\bar{\Psi}\right)e_{a}^{\mu}\gamma^{a}\Psi\right],\label{eq:q_action}
\end{equation}
where $\gamma^{a}$ are the 4 by 4 gamma matrices and $e^\mu_a$ is the inverse of $e_\mu^a$.
The action is written in this way to ensure Hermiticity locally.
The torsions or torsional electromagnetic fields are defined as $T_{\mu\nu}^{a}=\partial_{\mu}e_{\nu}^{a}-\partial_{\nu}e_{\mu}^{a}$.
%Hereafter we assume $u^{a}$ is small compared to the crystal constants.

%%%%%%%%%%%%%%%%%%%%%%%%%%%%%%%%%%%%%
\section{Band-projected Green's function and Wigner's transformation}\label{Green}
%\textcolor{blue}{In this section, we should project the two-band Green's function onto its positive-energy band. Then, by using the Wigner transformation, the Berry connection naturally arises.}

The Green's function for the right-handed Weyl fermions can be read
off from the action above directly. One can see that this Green's
function depends on both momentum and position. It is well-known that
in quantum physics, the position operator and the momentum operator
do not commute with each other. To develop a semiclassical theory
described by both momentum and position, one needs to utilize the
Wigner transformation and has the Wigner-transformed Green's function~(see Appendix. \ref{WigT})
\begin{equation}
i\tilde{G}^{-1}=\left(1+w\right)p_{\mu}\delta_{a}^{\mu}\sigma^{a}-w_{a}^{\mu}p_{\mu}\sigma^{a}+\mathcal{O}\left(u^{2}\right),\label{eq:green-1}
\end{equation}
where $\sigma^{a}=\left(1,\thinspace\sigma^{i}\right)$, $w_{a}^{\mu}\equiv\delta_{a}^{\rho}\partial_{\rho}u^{b}\delta_{b}^{\mu}$
and $w\equiv\delta_{\mu}^{a}w_{a}^{\mu}$. The factor $\left(1+w\right)$
comes from the determinant $\left|\det e_{\mu}^{a}\right|$. One can
see that, up to the linear-order terms in $u$, this Green's function
is equivalent to $i\tilde{G}^{-1}\approx\left(1+w\right)\left[p_{\mu}\delta_{a}^{\mu}-\left(p_{\mu}w_{a}^{\mu}\right)\right]\sigma^{a}$.
Thus, $w_{a}^{\mu}$ couples to Weyl fermions with a coupling charge
$p$ in a way similar to the electromagnetic gauge fields.

In order to derive the semiclassical action, one shall project the
two-band Green's function in Eq. (\ref{eq:green-1}) onto its positive-energy
bands, i.e. $\tilde{G}_{++}^{-1}=\langle u_{+}|*\left(\tilde{G}^{-1}\right)*|u_{+}\rangle,$
where $|u_{+}\rangle$ is the positive-energy eigenstates: $\mathbf{p}\cdot\mathbf{\sigma}|u_{+}\rangle=\left|\mathbf{p}\right||u_{+}\rangle$~\cite{SCRegion}.
The Moyal star product, $*=\exp\left[-\frac{i}{2}\left(\overleftarrow{\partial}_{q^{\mu}}\overrightarrow{\partial}_{p_{\mu}}-\overleftarrow{\partial}_{p_{\mu}}\overrightarrow{\partial}_{q^{\mu}}\right)\right]$,
is from the Wigner transformation. After lengthy calculations, one gets the projected Green's function~(see Appendix. \ref{WigT})
\begin{equation}
i\tilde{G}_{++}^{-1}=i\mathcal{G}^{-1}-\mathbf{a}_{p}\cdot\partial_{\mathbf{q}}i\mathcal{G}^{-1}+\xi_{\text{viel}},\label{eq:green}
\end{equation}
where $i\mathcal{G}^{-1}=\left(1+w\right)p_{a}\delta_{\nu}^{a}\hat{p}^{\nu}+p_{a}w_{\nu}^{a}\hat{p}^{\nu}$ and
$a_{p}^{i}=\langle u_{+}|i\partial_{p_{i}}|u_{+}\rangle$ is the Berry
connection for electrons in the conduction band. The bold alphabet
here is used for vectors in Euclidean space, e.g. $q_{\mu}=\left(q^{0},\thinspace-\mathbf{q}^{i}\right)$
and $\partial_{\mathbf{q}}^{i}=\frac{\partial}{\partial\mathbf{q}^{i}}$
is the derivative with respect to coordinates $\mathbf{q}$. In Eq.
(\ref{eq:green}), the first two terms can be regarded as first-order
Taylor's expansion of $\mathcal{G}^{-1}\left(\mathbf{q}-\mathbf{a}_{p}\right)$.
Hence, compared to electromagnetic fields, the Berry connection is
like gauge fields in the momentum space. In addition, $\mathcal{G}^{-1}$
is the next-lowest-order expansion of $\left(\det e_{\mu}^{a}\right)p_{a}e_{\mu}^{a}\hat{p}^{\mu}$.
$\hat{p}^{\mu}$ originates from $e_{b}^{\mu}\langle u_{+}|\sigma^{b}|u_{+}\rangle$,
where $e_{b}^{\mu}$ links the locally flat spacetime to the lab coordinates.
Because, for right-handed Weyl fermions, the velocity operator is
$v^{a}=\partial H/\partial p_{a}=\sigma^{a}$, $\hat{p}^{\mu}$ would link to velocity in lab coordinate as we shall show later.
Finally, the energy correction $\xi_{\text{viel}}=\epsilon^{0\alpha\beta\sigma}\frac{\hat{p}_{\sigma}}{2\left|\mathbf{p}\right|}\left(\frac{1}{2}p_{b}T_{\alpha\beta}^{b}\right)$
describes the coupling between the orbital magnetic moment $\hat{\mathbf{p}}/2\left|\mathbf{p}\right|$
and the spatial components of the torsion tensor $T_{ij}^{a}$. The torsional
magnetic field $\tilde{\mathbf{T}}^{a}$ is defined as the Hodge dual
of the torsion tensor $\left(\mathbf{T}^{a}\right)^{jk}=\partial_{\mathbf{q}}^{j}\left(\mathbf{e}^{a}\right)^{k}-\partial_{\mathbf{q}}^{k}\left(\mathbf{e}^{a}\right)^{j}$,
i.e. $\left(\tilde{\mathbf{T}}^{a}\right)^{i}=\frac{1}{2}\epsilon^{ijk}\left(\mathbf{T}^{a}\right)^{jk}$.
%
%It should be noted that, since Weyl fermions are massless, the non-diagonal components of Green's function ($\tilde{G}_{+-}$ and $\tilde{G}_{-+}$)
%are generally not small.
%In this paper, we would like to focus on the semiclassical region ($\left|\mathbf{p}\right|\gg\sqrt{\left|\mathbf{B}\right|}$),
%in which the Fermi level crosses many Landau levels such that $\tilde{G}_{\pm\mp}$
%becomes negligible \citep{stephanov2012prl}.

%%%%%%%%%%%%%%%%%%%%%%%%%%%%%%%%%%%%%
\section{Semiclassical action and equations of motion \label{action and EOM}}
In this section, we would like to construct the semiclassical dynamics based on the band-projected Green function above.
The dispersion relation for the positive-energy particles can be obtained
by solving the equation $\tilde{G}_{++}^{-1}=0$. That is, the on-shell
particles are located at poles of Green's function. By keeping terms up
to order $u$ and restoring the electromagnetic fields, one can straightforwardly
find the solution to Eq. (\ref{eq:green}), leading to the semiclassical action~(see Appendix. \ref{WigT})
\begin{equation}
L=\mathbf{k}\cdot\dot{\mathbf{q}}-\left(\left|\mathbf{k}\right|-\xi_{\text{viel}}-\xi_{\text{em}}\right)+\left[\left(w^{a}\right)_{\mu}k_{a}-A_{\mu}\right]\dot{q}^{\mu}-\mathbf{a}_{k}\cdot\dot{\mathbf{k}},\label{Lag1}
\end{equation}
where $k_{a}=\left(\left|\mathbf{k}\right|,\thinspace-\mathbf{k}^{i}\right)$,
$\dot{q}^{\mu}=\left(1,\thinspace\dot{\mathbf{q}}^{i}\right)$, $A_{\mu}=\left(\phi,\thinspace-\mathbf{A}^{i}\right)$
is the electromagnetic gauge potential and $\xi_{\text{em}}=\epsilon^{0\alpha\beta\sigma}\frac{1}{2\left|\mathbf{p}\right|}\left(\frac{1}{2}F_{\alpha\beta}\right)\hat{p}_{\sigma}$
stems from the orbital magnetic moment of electrons. It is clear that
dislocations modify the semiclassical action through two ways: the
shift of the gauge potential and the correction of the energy dispersion,
which implies that $\left(w^{a}\right)_{\mu}$ does behave like the
electromagnetic gauge fields $A_{\mu}$ but with a coupling charge
$k_{a}$. Note that we have changed variable from canonical momentum
$\mathbf{p}$ to mechanical momentum $\mathbf{k}=\mathbf{p}+\left(\mathbf{w}^{0}\right)\left|\mathbf{p}\right|+\left(\mathbf{w}^{j}\right)p_{j}-\mathbf{A}$.

The corresponding equations of motion can be derived from the semiclassical
action in Eq. (\ref{Lag1}),
\begin{eqnarray}
D\dot{\mathbf{q}}^{i} & = & \left\{ \left[1-\partial_{\mathbf{k}}^{n}\left(k_{a}\mathbf{w}^{a}\right)^{n}\right]\delta^{ij}+\partial_{\mathbf{k}}^{i}\left(k_{a}\mathbf{w}^{a}\right)^{j}\right\} \mathbf{v}^{j}\nonumber \\
 &  & -\epsilon^{ijk}\left[\tilde{\Omega}^{k}-\tilde{\Omega}^{m}\partial_{\mathbf{k}}^{m}\left(k_{a}\mathbf{w}^{a}\right)^{k}\right]\left(\partial_{\mathbf{q}}^{j}\mathcal{E}-\mathbf{T}_{\text{ele}}^{j}\right)\nonumber \\
 &  & -\left(\tilde{\Omega}\cdot\mathbf{v}\right)\mathbf{T}_{\text{mag}}^{i},\label{eq:eom_q}
\end{eqnarray}
and
\begin{eqnarray}
D\dot{\mathbf{k}}^{i} & = & -\left\{ \left[1-\partial_{\mathbf{k}}^{n}\left(k_{a}\mathbf{w}^{a}\right)^{n}\right]\delta^{ij}+\partial_{\mathbf{k}}^{j}\left(k_{a}\mathbf{w}^{a}\right)^{i}\right\} \left(\partial_{\mathbf{q}}^{j}\mathcal{E}-\mathbf{T}_{\text{ele}}^{j}\right)\nonumber \\
 &  & +\epsilon^{ijk}\mathbf{v}^{j}\left[-\delta^{km}+\partial_{\mathbf{k}}^{k}\left(k_{a}\mathbf{w}^{a}\right)^{m}\right]\mathbf{T}_{\text{mag}}^{m}\nonumber \\
 &  & +\tilde{\Omega}^{i}\left[\mathbf{T}_{\text{mag}}\cdot\left(\partial_{\mathbf{q}}\mathcal{E}-\mathbf{T}_{\text{ele}}\right)\right].\label{eq:eom_p}
\end{eqnarray}
where $\mathbf{v}^{i}=\partial_{\mathbf{k}}^{i}\left(\mathcal{E}-w_{0}^{a}k_{a}\right)$
is the velocity with $\partial_{\mathbf{k}}^{i}=\partial/\partial_{\mathbf{k}^{i}}$
being the derivative with respect to momentum $\mathbf{k}$. $\tilde{\mathbf{\Omega}}^{i}\equiv\frac{1}{2}\epsilon^{ijk}\Omega^{jk}$
is the Hodge dual of the Berry curvature $\Omega^{ij}\equiv\partial_{\mathbf{k}}^{i}\mathbf{a}_{k}^{j}-\partial_{\mathbf{k}}^{j}\mathbf{a}_{k}^{i}$.
In addition to the torsional magnetic fields, there also exists the
torsional electric fields $\left(\mathbf{T}_{0}^{a}\right)^{i}=-T_{0i}^{a}$,
which links to the thermal transport \citep{luttinger1964pr,qin2011prl,shitade2014ptep,tatara2015prl}.
Due to the common role played by torsions and the electromagnetic
fields, we could define $\mathbf{T}_{\text{mag}}^{i}=-\mathbf{B}^{i}+k_{a}\left(\tilde{\mathbf{T}}^{a}\right)^{i}$
and $\mathbf{T}_{\text{ele}}^{i}=\mathbf{E}^{i}+k_{a}\left(\mathbf{T}_{0}^{a}\right)^{i}$.
The modified density of states is given as $D/\left(2\pi\right)^{3}$
with $D=1-\tilde{\Omega}^{j}\mathbf{T}_{\text{mag}}^{j}+\tilde{\Omega}^{l}\partial_{\mathbf{k}}^{l}\left(k_{a}\mathbf{w}^{a}\right)^{i}\mathbf{T}_{\text{mag}}^{i}$.
With no torsions, $D$ would reduce to $1+\tilde{\mathbf{\Omega}}\cdot\mathbf{B}$,
which is well-known in semiclassical physics \citep{Culcer2004PRL,xiaoD2005prl,GaoY2014PRL,YangSY2015PRL,JiangQD2015PRL,DaiX2017PRL,niu2017Berry}.
Interestingly, the torsion coupling charge $k$ leads to an extra term in $D$.
Eqs. (\ref{Lag1})-(\ref{eq:eom_p}) are part of the main results in this work.

Let us now turn to the physics encoded in Eq. (\ref{eq:eom_q}). The
terms in the first line show that the velocity is modified by
torsions. The terms in the second line correspond to the anomalous
Hall effect. Because temperature gradient can be defined as $T_{0i}^{0}$
\citep{tatara2015prl} with $e_{\mu}^{0}$ coupling with the energy
current, the anomalous thermoelectric effect is also included. The term in the
last line contains both the CME and the torsional
CME~(see Appendix. \ref{TCME}). To be more specific, the current caused by
torsional magnetic fields is $-\int\frac{d^{3}k}{\left(2\pi\right)^{3}}f_{n}\left(\tilde{\Omega}_{n}\cdot\mathbf{v}_{n}\right)\left(\tilde{\mathbf{T}}^{a}k_{a}\right),$
where index $n$ denotes bands and chirality. $f_{n}$ is the Fermi-Dirac
distribution function. Because $f_{n}\left(\tilde{\Omega}_{n}\cdot\mathbf{v}_{n}\right)\left(\tilde{\mathbf{T}}^{i}k_{i}\right)$
is an odd function of momentum, this current should vanish unless
a pair of Weyl nodes with opposite chirality located at different
positions in energy-momentum space, which can be implemented through
breaking either time reversal symmetry or inversion symmetry.

Meanwhile, the terms in the first line of Eq. (\ref{eq:eom_p}) are
the electric force and the counterpart from torsions. Similarly, those
in the second line are the Lorentz force and its torsional counterpart.
One can clearly find that the torsional magnetic fields behave expectedly like the conventional
magnetic fields. The last term closely relates to the Liouville anomaly.

The anomaly term on the right-hand side in Eq. (\ref{eq:eom_p}) contains a mixing term between the electromagnetic
fields and torsions. But the Berry curvature always leads to a
Dirac delta function in the anomaly equation. Since the coupling charge of torsions
is momentum, non-trivial results require Weyl nodes deviate from the origin.
Hence, we assume that Weyl nodes with chirality $s$ ($s=\pm1$) are located at $\lambda_{sa}$, and then have
an extra term in Eq. (\ref{eq:q_action}), i.e. $\Sigma_{s}\int\left|\det e_{\mu}^{a}\right|\bar{\Psi}\left(-\lambda_{sa}P_{s}\right)\gamma^{a}\Psi$ with
$P_{s}=\frac{1}{2}\left(1+s\gamma^{5}\right)$ and $\lambda_{sa}=\left(\lambda_{s0},\thinspace-\boldsymbol{\lambda}_{s}^{i}\right)$, where  we have restored chirality index.
The corresponding semiclassical action becomes
\begin{eqnarray}
L_{s} & = & \mathbf{k}\cdot\dot{\mathbf{q}}-\left(\left|\mathbf{k}-\boldsymbol{\lambda}_{s}\right|+\lambda_{s0}-\xi_{\text{viel}}-\xi_{\text{em}}\right)\nonumber \\
 &  & +\left(w^{a}\right)_{\mu}k_{a}\dot{q}^{\mu}-A_{\mu}\dot{q}^{\mu}-\mathbf{a}_{ks}\cdot\dot{\mathbf{k}},\label{eq:action_semi}
\end{eqnarray}
where $\mathbf{a}_{ks}=\mathbf{a}_{ks}\left(\mathbf{k}-\mathbf{\lambda}_{s}\right)$
is a function of $\left(\mathbf{k}-\mathbf{\lambda}_{s}\right)$.
The energy correction from the orbital magnetic moment term becomes
$\xi_{\text{viel}}+\xi_{\text{em}}=\hbar\epsilon^{0\alpha\beta\sigma}\frac{\frac{1}{2}\left(k_{b}T_{\alpha\beta}^{b}+F_{\alpha\beta}\right)\widehat{k-\lambda_{s}}_{\sigma}}{2\left|\mathbf{k}-\boldsymbol{\lambda}_{s}\right|}$.
Note that, up to the lowest-order in external fields, the
dispersion relation is $k_{0}=\left|\mathbf{k}-\boldsymbol{\lambda}_{s}\right|+\lambda_{s0}$.

%%%%%%%%%%%%%%%%%%%%%%%%%%%%%%%%%%%%%
\section{Torsional responses and Liouville Anomaly}\label{responses_anomaly}
In this section, we turn to consider the torsional responses and derive the anomaly equations within the semiclassical formalism. We also predict a torsion modified AHE and discuss its experimental implementation. The Callan-Harvey mechanism is used to discuss the local charge conservation.

From the equation of motions in Eq. (\ref{eq:eom_q}), one finds the current stemming from $\left(\tilde{\Omega}\cdot\mathbf{v}\right)\mathbf{T}_{\text{mag}}^{a}k_{a}$ and $\xi_{\text{viel}}$ as

\begin{equation}
\mathbf{j}=\frac{\Lambda\lambda_{i}}{2\pi^{2}}\tilde{\mathbf{T}}^{i}+\frac{\mu\lambda_{i}}{2\pi^{2}}\left(1+\frac{1}{3}\right)\tilde{\mathbf{T}}^{i}+\frac{\mu}{\pi^{2}}\left[\mu_{5}\left(\frac{1}{2}+\frac{1}{3}\right)-\frac{\lambda_{0}}{6}\right]\tilde{\mathbf{T}}^{0},\label{eq:torsional_cme}
\end{equation}
where $\Lambda$ is the energy cut-off rather than the momentum cut-off
and is actually from the distribution function for negative-energy
particles: $f_{s-}=\left\{ \exp\left[\beta\left(-\left|\mathbf{k}\right|+\lambda_{s0}-\mu_{s}\right)\right]+1\right\} ^{-1}$.
The chemical potential for $s$-Weyl fermions is $\mu_{s}=\mu+s\mu_{5}$
and $\mu_{5}$ is the chiral chemical potential induced by the chiral
anomaly. In addition, we have assumed $\lambda_{s\mu}=s\lambda_{\mu}$ hereafter.

The first term on the right-hand side of Eq. (\ref{eq:torsional_cme}) is the torsional CME~(see Appendix. \ref{TCME}). The relevant current is proportional
to energy cut-off $\Lambda$, which is actually from the distribution
function for negative-energy particles. Note that the $\Lambda$-dependent
current had been tested numerically in a tight-binding model \citep{sumiyoshi2016prl}.
The coefficients of $1$ and $1/3$ in the second term come from $\left(\tilde{\Omega}\cdot\mathbf{v}\right)\mathbf{T}_{\text{mag}}^{i}$
and $\xi_{\text{viel}}$, respectively, which are first obtained in
the present work. The second term means that the torsional magnetic
fields can induce currents proportional to the chemical potential
rather than the chiral chemical potential. Physically, in the presence
of an external magnetic field, Weyl fermions with different chirality
would move oppositely and the net current is thus proportional to
the chiral chemical potential, which gives rise to the CME \citep{nielsen1983plb}.
On the other hand, for torsional magnetic fields, $\lambda_{si}$
provides an extra minus sign, so the current turns out to be proportional
to the chemical potential $\mu$. As we shall show later, the current
from $\left(\tilde{\Omega}\cdot\mathbf{v}\right)\mathbf{T}_{\text{mag}}^{i}$
is closely related to the Liouville anomaly in Eq. (\ref{eq:action_semi}).
Compared with the chiral pseudomagnetic effect \citep{zhou2013cpl,grushin2016prx,pikulin2016prx,huangprb2017},
both currents are proportional to the chemical potential. However,
the extra minus sign in the chiral pseudomagnetic effect comes from
the opposite coupling between the axial gauge fields and the right-handed
or left-handed Weyl fermions.
For the last term, the coefficient of $\mu_{5}/2$ comes from $\left(\tilde{\Omega}\cdot\mathbf{v}\right)\mathbf{T}_{\text{mag}}^{0}k_{0}$,
both the coefficients $\mu_{5}/3$ and $\lambda_{0}/6$ come from
$\xi_{\text{viel}}$. Because $\tilde{\mathbf{T}}^{0}$ links to the
background rotation, $\mu\mu_{5}\tilde{\mathbf{T}}^{0}/2\pi^{2}$
corresponds to the chiral vortical effect \citep{vilenkin1979prd,vilenkin1980prd,stephanov2012prl}.
In analogy to the dynamical CME \citep{KharzeevPRD2009,ma2015prb,zhong2016prl,ZhouJH2018PRB},
$-\lambda_{0}\mu\tilde{\mathbf{T}}^{0}/6\pi^{2}$ can be regarded as the dynamical chiral
vortical effect, which stems from the orbital magnetic moment of electrons
on the Fermi surface as well.

The Liouville theorem states that the phase-space volume does not change under evolution.
If we define $\Omega_{L}$ as the volume form in the extended phase space (position,
momentum and time), then the Liouville theorem is equivalent to $\mathcal{L}_{V}\Omega_{L}=0$.
$V=\dot{q}^{i}\frac{\partial}{\partial q^{i}}+\dot{k}_{i}\frac{\partial}{\partial k_{i}}+\frac{\partial}{\partial t}$
is a vector relates to translation along time: for an arbitrary function
$g\left(q,\thinspace k,\thinspace t\right)$, $Vg=\frac{d}{dt}g$.
$\mathcal{L}_{V}$ is the Lie derivative along vector $V$. Then, in
the presence of the Berry connection, it can be shown that $\mathcal{L}_{V}\Omega_{L}\propto d\Omega\times\left(\dots\right)$
\citep{dwivedi2013iop}, where $\Omega^{jk}$ is the Berry curvature,
$d\Omega=\partial_{k_{i}}\Omega^{jk}dk_{i}\wedge dk_{j}\wedge dk_{k}$.
That is, the Liouville equation no longer holds because of singularities of the Berry curvature at the Weyl nodes.
Hence, the Liouville anomaly originates from the infrared physics.
But the Nieh-Yan term states that $\partial_{\mu}j^{5\mu}=\frac{\Lambda_{r}^{2}}{16\pi^{2}}\epsilon^{\mu\nu\rho\sigma}\partial_{\mu}e_{\nu}^{a}\partial_{\rho}e_{\sigma}^{b}$,
where $\Lambda_{r}$ is the energy-momentum cut-off. So it is aware
of cut-off and thus conflicts with the picture from semiclassical
physics. From the Liouville equation in the collisionless limit,
one reaches the anomaly equation in the presence of both the electromagnetic
fields and torsions ~(see Appendix. \ref{LiouA})
\begin{equation}
\partial_{\mu}j_{s}^{\mu}=-\frac{s\epsilon^{\mu\nu\rho\sigma}}{32\pi^{2}}\left(F_{\mu\nu}F_{\rho\sigma}+\lambda_{sa}\lambda_{sb}T_{\mu\nu}^{a}T_{\rho\sigma}^{b}-2\lambda_{sa}F_{\mu\nu}T_{\rho\sigma}^{a}\right).\label{eq:anomaly}
\end{equation}
where $j_{s}^{\mu}=\left(j_{s}^{0},\mathbf{j}_{s}\right)$ is the current of Weyl fermions with chirality $s$.
This new Liouville anomaly is another main result in our work.
It is clear that the last two terms explicitly depend on the positions of Weyl nodes in energy-momentum space and thus significantly differ from the counterparts of axial gauge fields~\cite{huangprb2017}.
Although these axial gauge fields from crystal deformations do not change our main results, they may become significant out of the weak displacement field regime in real materials.
According to Ref.~\citep{xiao2006prl}, the coupling charge between the torsional electric
fields (or temperature gradient) and particles is $k_{0}-\mu_{s}$
rather than $k_{0}$, where terms proportional to temperature are
neglected for simplicity. This shift of the coupling charge leads
to some extra terms in Eq. (\ref{eq:anomaly}), one of which proportional
to $\frac{-s\mu_{s}}{4\pi^{2}}\left[\left(\mathbf{T}_{0}^{0}\right)^{i}\mathbf{B}^{i}\right]$
can provide an intuitive explanation to the recent negative
magnetothermal resistance in the Weyl semimetal NbP \citep{gooth2017nature}.

It is straightforward to derive the axial current
\begin{equation}
\partial_{\mu}j^{5\mu}=-\frac{\epsilon^{\mu\nu\rho\sigma}}{16\pi^{2}}\left(F_{\mu\nu}F_{\rho\sigma}+\lambda_{a}\lambda_{b}T_{\mu\nu}^{a}T_{\rho\sigma}^{b}\right),\label{anomaly_ja}
\end{equation}
and the continuity equations for Weyl fermions,
\begin{equation}
\partial_{\mu}j^{\mu}=\frac{\epsilon^{\mu\nu\rho\sigma}}{8\pi^{2}}\left(\lambda_{a}F_{\mu\nu}T_{\rho\sigma}^{a}\right),\label{anomaly_jb}
\end{equation}
which can be understood from the chiral zeroth Landau level~(see Appendix.~\ref{LLs}). Assuming the screw dislocations along the $z$-axis, the Weyl nodes with chirality $s$ locate at $s\lambda_{\mu}$ with $s=\pm1$. For simplicity, we set $\lambda_{\mu}=(0,\,0,\,0,\,-\lambda_{z})$ and $\lambda_{z}>0$.
The displacement vector is assumed to be along the $z$-axis, so only one component of the vielbeins survives $e^{3}_{\mu}=\frac{1}{2}(0,\,-\tilde{T}y,\,\tilde{T}x,\,0)$.
Note that the surface density of the Burgers vector fields rather than itself is constant. Consequently, the zeroth Landau levels near $p_{z}=s\lambda_{z}$ are $p_z-s\lambda_{z}$. Turning on an electric field along the $z$-axis, charges are pumped up from the Dirac sea and extra particles are "produced". Specifically, the level degeneracy is roughly $\frac{\lambda_{z}\tilde{T}}{2\pi}$ in the vicinity of $p_{z}=s\lambda_z$ and the variation of momentum is, $\triangle p_z=E\triangle t$. Hence, the total variation of charge density is given as $\partial_{t} j^0=\frac{1}{2\pi^2}\lambda_{z}\tilde{T}E$, whose covariant form is Eq.~(11).

The first term in Eq. $\eqref{anomaly_ja}$ corresponds to the conventional
chiral anomaly \citep{adler1969pr,bell1969nca}. Unlike the Nieh-Yan
term, the second term specifically depends on the locations of Weyl
nodes but is independent of the cut-off. A finite chiral chemical
potential could be developed by a time dependent dislocation even
without any external electromagnetic fields and would be crucial to the anomalous transport phenomena for Weyl semimetals~\cite{gorbar2018ltp}.
\begin{figure}
\includegraphics[scale=0.5]{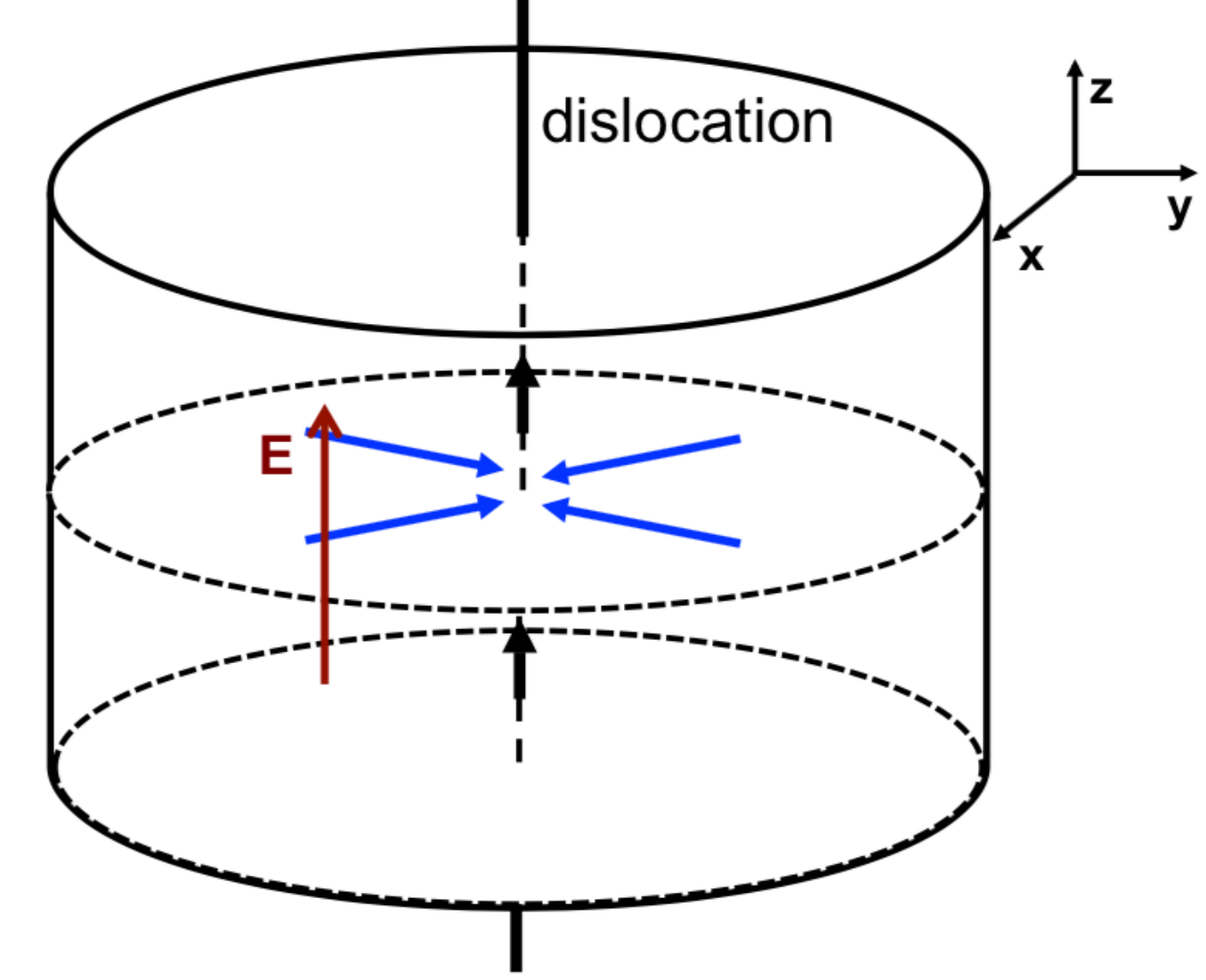}\caption{Schematic picture of the Callan-Harvey mechanism for the cancelation of the anomaly due to the chiral zero modes in the dislocation
by the bulk states through the torsion modified AHE.
The dashed vertical line refers to the dislocation along the $z$ direction. The black arrows refer to the chiral anomalous current along the dislocation and blue arrows to the torsion modified anomalous Hall current under an external electric field along the $z$ direction.~\label{Figchm}}
\end{figure}

From the anomaly equation in Eq.~(\ref{anomaly_jb}), one finds the following solution~\cite{FootN1}
\begin{equation}
\mathbf{j}=-\frac{\lambda_{a}}{2\pi^{2}}\left(\mathbf{w}^{a}\times\mathbf{E}-w_{0}^{a}\mathbf{B}\right).\label{eq:tm_ahe}
\end{equation}
In this work, we mainly focus on the static dislocations and would
like to neglect the term proportional to $w_{0}^{a}$. $\mathbf{j}=-\frac{\lambda_{a}}{2\pi^{2}}\mathbf{w}^{a}\times\mathbf{E}$
is first discovered here and can be dubbed as the torsion modified AHE.
This resulting anomalous Hall current is still perpendicular
to the electric field, but can be parallel to the momentum spacing between
the two Weyl nodes with opposite chirality $\boldsymbol{\lambda}$.
To show this point, let us set both screw dislocations and $\boldsymbol{\lambda}$
along the $z$-axis, i.e. $u^{a}=\left(0,\thinspace0,\thinspace0,\thinspace\mathbf{u}^{3}\left(x,\thinspace y\right)\right)$.
In addition, the only non-vanishing $\mathbf{w}^{a}$ is $\mathbf{w}^{3}=\left(\left(\mathbf{w}^{3}\right)^{1},\thinspace\left(\mathbf{w}^{3}\right)^{2},\thinspace0\right).$
When the electric field is along the $x$-axis, the responses current
is along the $z$-axis, i.e. $j^{3}=\frac{\lambda_{3}}{2\pi^{2}}\left(\mathbf{w}^{3}\right)^{2}\mathbf{E}^{1}$.
One can see that the ratio between the torsion modified AHE and the intrinsic AHE is roughly equal to $|u|/a$ with $a$ being the crystal constant. For the Weyl semimetal Co$_{3}$Sn$_{2}$S$_{2}$, the giant intrinsic AHE is about $10^3~\Omega^{-1}\text{cm}^{-1}$~\cite{CoSnSLiu}. The torsion modified AHE can thus reach tens of $\Omega^{-1}\text{cm}^{-1}$ ($|u|/a$ is from 0.01 to 0.1.).
Note that the mixing term in Eq. (\ref{anomaly_jb}) also underlies
the terms of $\frac{\mu\lambda_{i}}{2\pi^{2}}\tilde{\mathbf{T}}^{i}$
in Eq. (\ref{eq:torsional_cme}). In the thermal field theory, the
chemical potential would couple to $j^{0}$, i.e. $\int\mu j^{0}$.
So by keeping terms to leading order in $\mu$, the effective action from this
mixing term is $\frac{\epsilon^{0ijk}}{4\pi^{2}}\int\mu A_{i}\left(\lambda_{a}T_{jk}^{a}\right)$.
Thus, the response current is given as $\mathbf{j}=\frac{\mu}{2\pi^{2}}\lambda_{i}\tilde{\mathbf{T}}^{i}$~\cite{FootN2}.

Interestingly, Eq. (\ref{anomaly_jb}) involves a mixing term between the electromagnetic fields and torsions,
which seems to violate the charge conservation.
However, one can understand this anomaly equation from the Callan-Harvey mechanism~\cite{callan1985npb}.
Namely, anomalies due to the chiral zero modes localized in defects are cancelled by the compensation of those states from the bulk.
The gauge invariance is thus locally preserved, leading to the local conservation of electric charges.
It has been shown that, in Weyl semimetals with a dislocation, there are the chiral zero modes with opposite chirality localized in the dislocation and the boundary, respectively~\cite{tankane2017jpsj}. In addition, the chiral zero mode in the dislocation or anti-dislocation was also numerically found in Weyl semimetals~\citep{sumiyoshi2016prl}.
To specifically demonstrate the cancelation of the anomaly due to the chiral zero mode in the dislocation by the bulk states through the torsion modified AHE,
we consider a dislocation along the $z$-direction locates at $x=y=0$ (see Fig.~\ref{Figchm}).
Because of the chiral zero modes localized at the dislocation (black arrows), under external electric fields along the $z$-direction ($\mathbf{E}^3$), particles are pumped up from the Dirac sea, which implies a one-dimensional anomaly $\partial_\mu j^{\mu}\sim \frac{1}{2\pi}\mathbf{E}^3$ in the dislocation.
Note that  $(\mathbf{w}^a)^{1}=-\frac{b^a}{2\pi}\frac{y}{(x^2+y^2)}$ and $(\mathbf{w}^a)^{2}=+\frac{b^a}{2\pi}\frac{x}{(x^2+y^2)}$ are solutions to the equation $T^a_{\mu\nu}=-b^a \delta{(x)}\delta{(y)}$.
So the corresponding torsion modified anomalous Hall currents are $\mathbf{j}^{1}=-\frac{\lambda_{a}b^{a}}{2\pi^2}\frac{x}{(x^2+y^2)}\mathbf{E}^{3}$ and $\mathbf{j}^{2}=-\frac{\lambda_{a}b^{a}}{2\pi^2}\frac{y}{(x^2+y^2)}\mathbf{E}^{3}$ (blue arrows),
which means that this current flows toward (outward) the dislocation for $\lambda_{a}b^{a}>0$ ($\lambda_{a}b^{a}<0$).
Consequently, the chirality of localized chiral zero modes should depend on the sign of $\lambda_{a}b^{a}$ and
the extra charges in the dislocation are compensated by the bulk states.
Therefore, the electric charges are locally conserved.
Note that the torsion modified AHE here plays a similar role as the AHE in the cancellation of the anomaly of one-dimensional domain wall embedded in the (2+1)-dimensional massive Dirac fermion system~\cite{stone1991aop}.
%

%%%%%%%%%%%%%%%%%%%%%%%%%%%%%%%%%%%%%
\section{Conclusions and discussions}\label{conclusion}
In summary, we have presented a formalism to construct the semiclassical action and equations of motion for
Weyl fermions in the presence of both electromagnetic fields and torsions from the quantum field theory.
It has been shown that the torsional electromagnetic fields make the Liouville anomaly equation essentially different from the counterpart of axial gauge fields.
Our results could give rise to various torsional responses and reproduce the torsional CME and the chiral vortical effect.
In addition, a new torsion modified AHE originating from the mixing term in the Liouville anomaly is predicted and its implementation
in Weyl semimetals lacking time reversal symmetry is discussed as well.

Recent progress on material realization of Weyl semimetals with no time reversal symmetry in Co$_{3}$Sn$_{2}$S$_{2}$ could facilitate the experimental investigation of the torsional responses associated with dislocations.

\section*{acknowledgments}\label{ackledg}
\begin{acknowledgments}
The authors are grateful to Professor Michael Stone for pointing out the Callan-Harvey Mechanism and to Liang Dong, Bo Han, Tao Qin and Cenke Xu for insightful discussions.
ZMH, LL and HHZ were supported by the National Natural Science Foundation of China (NSFC) under Grant No. 11875327.
JZ was supported by the 100 Talents Program of Chinese Academy of Sciences.
\end{acknowledgments}

\begin{appendix}
\begin{widetext}

%%%%%%%%%%%%%%%%%%%%%%%%%%%%%%%%%%%%%%%%%%%%%%%%%%%%%%%%%%%%
\section{Wigner transformation, band projection and semiclassical action}\label{WigT}

In this section, we provide the detailed derivations of the Wigner-transformed and band-projected Green's function.
For clarity, we focus on the case in which the torsional electromagnetic fields and the electromagnetic
fields are parallel.

The action for Weyl semimetals in the presence of dislocations is
given as \cite{taylor2011prl,taylor2013prd}
\begin{equation}
S=\int d^{4}x\sqrt{-\det g}\frac{1}{2}\left[\bar{\Psi}e_{a}^{\mu}\gamma^{a}\left(i\partial_{\mu}\Psi\right)-\left(i\partial_{\mu}\bar{\Psi}\right)e_{a}^{\mu}\gamma^{a}\Psi\right],
\end{equation}
where $e_{a}^{\mu}$ is the frame fields, $\eta_{ab}$ is flat spacetime
coordinates with indices $a,\thinspace b=0,\thinspace1,\thinspace2,\thinspace3$
and $g_{\mu\nu}=e_{\mu}^{a}e_{\nu}^{b}\eta_{ab}$ is curved spacetime
metric with indices $\mu,\thinspace\nu=0,\thinspace1,\thinspace2,\thinspace3$.
The torsion is defined as $T_{\mu\nu}^{a}=\partial_{\mu}e_{\nu}^{a}-\partial_{\nu}e_{\mu}^{a}$.
If the lattice displacement is $u^{a}\left(x\right)$, then $e_{\mu}^{a}=\delta_{\mu}^{a}+\partial_{\mu}u^{a}$
and $e_{a}^{\mu}=\delta_{a}^{\mu}-\delta_{a}^{\rho}\partial_{\rho}u^{b}\delta_{b}^{\mu}+\mathcal{O}\left(u^{2}\right)$.

Then, by keeping terms up to order $\mathcal{O}\left(u\right)$, the
action for right-handed Weyl fermions becomes
\begin{equation}
S_{R}=\int d^{d}x\left\{ \Psi_{R}^{\dagger}\left[\sigma^{a}\delta_{a}^{\mu}i\partial_{\mu}+w\sigma^{\mu}\left(i\partial_{\mu}\right)+\frac{1}{2}i\left(\partial_{\mu}w\right)\sigma^{\mu}-w_{a}^{\mu}\sigma^{a}i\partial_{\mu}-\frac{1}{2}\left(i\partial_{\mu}w_{a}^{\mu}\right)\sigma^{a}\right]\Psi_{R}\right\}
\end{equation}
where $w_{a}^{\mu}=\delta_{a}^{\nu}\delta_{b}^{\mu}\partial_{\nu}u^{b}$
and $w=\delta_{a}^{\mu}\partial_{\mu}u^{a}$. The corresponding Green's
function is
\begin{equation}
iG^{-1}=\sigma^{a}\delta_{a}^{\mu}i\partial_{\mu}+\left[w_{b}^{\alpha}\left(i\sigma^{\mu}\partial_{\nu}\right)+\frac{1}{2}\partial_{\nu}w_{b}^{\alpha}i\sigma^{\mu}\right]\left(\eta_{\alpha}^{b}\eta_{\mu}^{\nu}-\eta_{\alpha}^{\nu}\eta_{\mu}^{b}\right).
\end{equation}
Performing the Wigner transformation \cite{rammer2007cambridge} leads to
\begin{eqnarray}
i\tilde{G}^{-1} & = & p_{\mu}\sigma^{a}\delta_{a}^{\mu}+\left(\eta_{\alpha}^{b}\eta_{\mu}^{\nu}-\eta_{\alpha}^{\nu}\eta_{\mu}^{b}\right)\left[w_{b}^{\alpha}*p_{\nu}\sigma^{\mu}+i\frac{1}{2}\partial_{\nu}w_{b}^{\alpha}\sigma^{\mu}\right]\nonumber \\
 & = & p_{\mu}\sigma^{a}\delta_{a}^{\mu}+\left(\eta_{\alpha}^{b}\eta_{\mu}^{\nu}-\eta_{\alpha}^{\nu}\eta_{\mu}^{b}\right)\left(w_{b}^{\alpha}p_{\nu}\sigma^{\mu}\right),
\end{eqnarray}
where $*=\exp\left[-\frac{i}{2}\hbar\left(\overleftarrow{\partial}_{q^{\mu}}\overrightarrow{\partial}_{p_{\mu}}-\overleftarrow{\partial}_{p_{\mu}}\overrightarrow{\partial}_{q^{\mu}}\right)\right]$
is Moyal's product from the Wigner's transformation. Although we have
used the nature unit, for heuristic purposes, $\hbar$ in Moyal's
product will not be set to $1$ hereafter. Because we are most interested
in semiclassical limit, keeping $\hbar$ in Moyal's product enables us to keep track of this.

In addition, we project $\tilde{G}$ onto the space spanning by its
positive eigenstates to obtain semiclassical action, e.g. $|u_{\pm}\rangle$
with $\mathbf{p}\cdot\mathbf{\sigma}|u_{\pm}\rangle=\pm\left|\mathbf{p}\right||u_{\pm}\rangle$.
For simplicity, we focus on the positive-energy band and the counterpart
for the negative-energy band is similar. The projected Green's function becomes
\begin{eqnarray}
i\tilde{G}_{++}^{-1} & = & \langle u_{+}|*\tilde{G}^{-1}*|u_{+}\rangle\nonumber \\
 & = & \langle u_{+}|\left[1-\frac{i\hbar}{2}\left(\overleftarrow{\partial}_{q^{\mu}}\overrightarrow{\partial}_{p_{\mu}}-\overleftarrow{\partial}_{p_{\mu}}\overrightarrow{\partial}_{q^{\mu}}\right)\right]\left[p_{\mu}\sigma^{a}\delta_{a}^{\mu}+\left(\eta_{\alpha}^{b}\eta_{\mu}^{\nu}-\eta_{\alpha}^{\nu}\eta_{\mu}^{b}\right)\left(w_{b}^{\alpha}p_{\nu}\sigma^{\mu}\right)\right]\nonumber \\
 &  & \times\left[1-\frac{i\hbar}{2}\left(\overleftarrow{\partial}_{q^{\mu}}\overrightarrow{\partial}_{p_{\mu}}-\overleftarrow{\partial}_{p_{\mu}}\overrightarrow{\partial}_{q^{\mu}}\right)\right]|u_{+}\rangle+\mathcal{O}\left(\hbar^{2}\right)\nonumber \\
 & = & \left(p_{0}-\left|\mathbf{p}\right|\right)\left(1+w\right)-\left[w_{0}^{0}p_{0}+\left(w_{i}^{0}p_{0}\hat{p}^{i}+w_{0}^{i}p_{i}\right)+w_{i}^{j}p_{j}\hat{p}^{i}\right]\nonumber \\
 &  & +\left[p_{0}\left(\partial_{q^{\mu}}w\right)-\partial_{q^{\mu}}w_{0}^{0}p_{0}-\left(\partial_{q^{\mu}}w_{0}^{i}\right)p_{i}\right]\left\{ \frac{i}{2}\hbar\left[\left(\partial_{p_{\mu}}\langle u_{+}|\right)|u_{+}\rangle-\langle u_{+}|\left(\partial_{p_{\mu}}|u_{+}\rangle\right)\right]\right\} \nonumber \\
 &  & +\left[-p^{i}\left(\partial_{q^{\mu}}w\right)-\left(\partial_{q^{\mu}}w_{i}^{0}\right)p_{0}-\left(\partial_{q^{\mu}}w_{i}^{j}\right)p_{j}\right]\left\{ \frac{i}{2}\hbar\left[\left(\partial_{p_{\mu}}\langle u_{+}|\right)\sigma^{i}|u_{+}\rangle-\langle u_{+}|\sigma^{i}\left(\partial_{p_{\mu}}|u_{+}\rangle\right)\right]\right\} +\mathcal{O}\left(\hbar^{2}\right).
\end{eqnarray}
where $p_{\mu}=\left(p_{0},\thinspace-\mathbf{p}\right)$. We have
only keep terms up to order $\hbar$ in the second line. In the fourth
line, we have used $\langle u_{+}|\mathbf{\sigma}^{i}|u_{+}\rangle=\hat{p}^{i}$.
The Berry connection is defined as
\begin{eqnarray}
\langle u_{+}\left(p\right)|\partial_{p_{\mu}}|u_{+}\left(p\right)\rangle=-i\mathcal{A}_{p}^{\mu}=-i\left(0,\thinspace\mathbf{a}_{p}\right).
\end{eqnarray}
Because
\begin{eqnarray}
\langle u_{+}\left(p\right)|\sigma^{i}\partial_{p_{\mu}}|u_{+}\left(p\right)\rangle=\langle u_{+}\left(p\right)|\sigma^{i}\left(\frac{|u_{+}\left(p\right)\rangle-|u_{+}\left(p-\triangle p\right)\rangle}{\triangle p_{\mu}}\right),
\end{eqnarray}
by use of the modified Gordon's identity in Appendix. \ref{MGI}, one finds
\begin{eqnarray}
\text{Im}\langle u_{+}|\sigma^{i}\left(\partial_{p_{\mu}}|u_{+}\rangle\right)=-i\mathcal{A}_{p}^{\mu}\hat{p}^{i}+i\epsilon^{i\mu k}\frac{\hat{p}_{k}}{2\left|\mathbf{p}\right|}|_{\mu\neq0}+\mathcal{O}\left(\triangle p\right).
\end{eqnarray}
and
\begin{eqnarray}
\text{Im}\left(\partial_{p_{\mu}}\langle u_{+}|\right)\sigma^{i}|u_{+}\rangle=i\mathcal{A}_{p}^{\mu}\hat{p}^{i}-i\epsilon^{i\mu k}\frac{\hat{p}_{k}}{2\left|\mathbf{p}\right|}|_{\mu\neq0}+\mathcal{O}\left(\triangle p\right).
\end{eqnarray}
Thus, the projected Green's function becomes
\begin{eqnarray}
i\tilde{G}_{++}^{-1} & = & i\mathcal{G}_{0}^{-1}\left(q\right)-\hbar\mathcal{A}_{p}^{\mu}\partial_{q^{\mu}}i\mathcal{G}_{0}^{-1}-\hbar\epsilon^{imk}\frac{\hat{p}_{k}}{2\left|\mathbf{p}\right|}\left[\left(\partial_{q^{m}}w_{i}^{0}p_{0}+\partial_{q^{m}}w_{i}^{j}p_{j}\right)\right],
\end{eqnarray}
where $i\mathcal{G}_{0}^{-1}\left(q\right)=\left(p_{0}-\left|\mathbf{p}\right|\right)\left(1+w\right)-w_{a}^{\mu}p_{\mu}\hat{p}^{a}$
and $\hat{p}^{a}=\left(1,\thinspace\hat{\mathbf{p}}\right)$. Because
$\hat{p}^{a}$ is from $\langle u_{+}|\sigma^{a}|u_{+}\rangle$, it
is supposed to link to velocity. In reality, the response current
is measured in lab coordinates with index $\mu$, so we shall change
indices of $\hat{p}$ and this leads to
\begin{eqnarray}
i\tilde{G}_{++}^{-1}=i\mathcal{G}^{-1}\left(q\right)-\hbar\mathcal{A}_{p}^{\mu}\partial_{q^{\mu}}i\mathcal{G}^{-1}\left(q\right)+\xi_{\text{viel}},
\end{eqnarray}
with
\begin{eqnarray}
i\mathcal{G}^{-1}\left(q\right)=\left(1+w\right)p_{a}\delta_{\nu}^{a}\hat{p}^{\nu}+p_{a}w_{\nu}^{a}\hat{p}^{\nu},
\end{eqnarray}
and
\begin{eqnarray}
\xi_{\text{viel}}=\hbar\epsilon^{0\alpha\beta\sigma}\frac{\left(\frac{1}{2}p_{b}T_{\alpha\beta}^{b}\right)\hat{p}_{\sigma}}{2\left|\mathbf{p}\right|}.
\end{eqnarray}
Because
\begin{eqnarray}
\mathcal{G}^{-1}\left(q\right)-\hbar\mathcal{A}_{p}^{\mu}\partial_{q^{\mu}}\mathcal{G}^{-1}\left(q\right)\simeq\mathcal{G}^{-1}\left(q-\hbar\mathcal{A}_{p}^{\mu}\right)+\mathcal{O}\left(\hbar^{2}\right),
\end{eqnarray}
the Berry connection acts as gauge fields in momentum space. Then,
the dispersion relation can be determined by the poles of Green's function,
\begin{eqnarray}
p_{0} & \simeq & \left|\mathbf{p}\right|-w_{\mu}^{a}p_{a}\hat{p}^{\mu}-\xi_{\text{viel}}+\mathbf{a}_{p}\cdot\dot{\mathbf{p}}\nonumber \\
 & \simeq & \left|\mathbf{p}+\left(\mathbf{w}^{0}\right)\left|\mathbf{p}\right|+\left(\mathbf{w}^{j}\right)p_{j}\right|-\left[\left(w^{0}\right)_{0}\left|\mathbf{p}\right|+\left(w^{i}\right)_{0}p_{i}\right]-\xi_{\text{viel}}+\mathbf{a}_{p}\cdot\dot{\mathbf{p}},
\end{eqnarray}
where we have neglected terms of order $\mathcal{O}\left(u^{2}\right)$
and $\mathcal{O}\left(\hbar^{2}\right)$. After changing variable
from $\mathbf{p}$ to $\mathbf{k}=\mathbf{p}+\left(\mathbf{w}^{0}\right)\left|\mathbf{p}\right|+\left(\mathbf{w}^{j}\right)p_{j}$,
one gets the action
\begin{eqnarray}
L=\mathbf{k}\cdot\dot{\mathbf{q}}-\left(\left|\mathbf{k}\right|-\xi_{\text{viel}}\right)+w_{\mu}^{a}k_{a}\dot{q}^{\mu}-\mathbf{a}_{k}\cdot\dot{\mathbf{k}},
\end{eqnarray}
with $k_{\mu}=\left(\left|\mathbf{k}\right|,\thinspace-\mathbf{k}\right)$,
$\dot{q}^{\mu}=\left(1,\thinspace\dot{\mathbf{q}}\right)$ and
\begin{eqnarray}
\xi_{\text{viel}}=\hbar\epsilon^{0\alpha\beta\sigma}\frac{\left(\frac{1}{2}k_{b}T_{\alpha\beta}^{b}\right)\hat{k}_{\sigma}}{2\left|\mathbf{k}\right|}.
\end{eqnarray}

%%%%%%%%%%%%%%%%%%%%%%%%%%%%%%

%%%%%%%%%%%%%%%%%%%%%%%%%%%%%%%%%%%%%%%%%%%%%%%%%%%%%%%%%%%%
\section{Torsional chiral magnetic effect }\label{TCME}

In this section, both the torsional chiral magnetic effect and chiral
magnetic effect are derived from equations of motion with a careful
treatment of cut-off. For simplicity, we consider the zero-temperature limit.

The torsional chiral magnetic effect comes from $\left(\tilde{\Omega}\cdot\mathbf{v}\right)\tilde{\mathbf{T}}^{i}$
in the equations of motion. For $a=i$, it becomes
\begin{eqnarray}
\mathbf{j}_{s}^{\left(1\right)} & = & -\int\frac{d^{3}k}{\left(2\pi\right)^{3}}\left[f_{s}^{+}\left(\tilde{\Omega}_{s}^{+}\cdot\mathbf{v}^{+}\right)\left(\tilde{\mathbf{T}}^{i}k_{i}\right)+f_{s}^{-}\left(\tilde{\Omega}_{s}^{-}\cdot\mathbf{v}^{-}\right)\left(\tilde{\mathbf{T}}^{i}k_{i}\right)\right]\nonumber \\
 & = & -\int\frac{d^{3}k}{\left(2\pi\right)^{3}}\left\{ \frac{1}{\exp\left[\beta\left(\left|\mathbf{k}\right|+\lambda_{s0}-\mu_{s}\right)\right]+1}+\frac{1}{\exp\left[\beta\left(-\left|\mathbf{k}\right|+\lambda_{s0}-\mu_{s}\right)\right]+1}\right\} \left(\tilde{\Omega}_{s}\cdot\mathbf{v}\right)\left(\tilde{\mathbf{T}}^{i}\lambda_{si}\right)\nonumber \\
 & = & \frac{s}{4\pi^{2}}\left\{ \int_{\lambda_{s0}}^{\Lambda}d\epsilon_{+}\frac{1}{\exp\left[\beta\left(\epsilon_{+}-\mu_{s}\right)\right]+1}+\int_{-\lambda_{s0}}^{\Lambda}d\epsilon_{-}\left[1-\frac{1}{\exp\left[\beta\left(\epsilon_{-}+\mu_{s}\right)\right]+1}\right]\right\} \left(\tilde{\mathbf{T}}^{i}\lambda_{si}\right)\nonumber \\
 & = & \frac{s\left(\tilde{\mathbf{T}}^{i}\lambda_{si}\right)}{4\pi^{2}}\left(\mu_{s}+\Lambda\right),
\end{eqnarray}
where $\pm$ denotes positive- and negative-energy bands. In the second
line, we have shifted variables $\mathbf{k}$ to $\mathbf{k}+\boldsymbol{\lambda}_{s}$.
$\epsilon_{\pm}$ is defined as $\epsilon_{\pm}=\left|\mathbf{k}\right|\pm\lambda_{s0}$.
$\Lambda$ refers to cut-off for energy rather than momentum, i.e.
$\lambda_{s0}<\left|\mathbf{k}\right|+\lambda_{s0}<\Lambda$ and $-\Lambda<-\left|\mathbf{k}\right|+\lambda_{s0}<\lambda_{s0}$.
So the energy ranges from $-\Lambda$ to $\Lambda$. $f^{\pm}$ is
the distribution function for positive- and negative-energy particles,
respectively. Note that $\Lambda$ now plays the role of energy reference
\citep{landsteiner2014PRB}.

If $a=0$, this current is
\begin{eqnarray}
\mathbf{j}_{s}^{\left(2\right)} & = & -\int\frac{d^{3}k}{\left(2\pi\right)^{3}}\left[f_{s}^{+}\left(\tilde{\Omega}_{s}^{+}\cdot\mathbf{v}^{+}\right)\left(\tilde{\mathbf{T}}^{0}k_{0}^{+}\right)+f_{s}^{-}\left(\tilde{\Omega}_{s}^{-}\cdot\mathbf{v}^{-}\right)\left(\tilde{\mathbf{T}}^{0}k_{0}^{-}\right)\right]\nonumber \\
 & = & -\int\frac{d^{3}k}{\left(2\pi\right)^{3}}\left(\tilde{\Omega}_{s}\cdot\mathbf{v}\right)\left\{ \frac{1}{\exp\left[\beta\left(\left|\mathbf{k}-\boldsymbol{\lambda}_{s}\right|+\lambda_{s0}-\mu_{s}\right)\right]+1}\left(\tilde{\mathbf{T}}^{0}\epsilon_{+}\right)+\frac{1}{\exp\left[\beta\left(-\left|\mathbf{k}-\boldsymbol{\lambda}_{s}\right|-\lambda_{s0}-\mu_{s}\right)\right]+1}\left(-\tilde{\mathbf{T}}^{0}\epsilon_{-}\right)\right\} \nonumber \\
 & = & \frac{s}{4\pi^{2}}\left\{ \int_{\lambda_{s0}}^{\Lambda}d\epsilon_{+}\frac{\tilde{\mathbf{T}}^{0}\epsilon_{+}}{\exp\left[\beta\left(\epsilon_{+}-\mu_{s}\right)\right]+1}+\int_{-\lambda_{s0}}^{\Lambda}d\epsilon_{-}\left[1-\frac{1}{\exp\left[\beta\left(\epsilon_{-}+\mu_{s}\right)\right]+1}\right]\left(-\tilde{\mathbf{T}}^{0}\epsilon_{-}\right)\right\} \nonumber \\
 & = & \frac{s}{8\pi^{2}}\left(\mu_{s}^{2}-\Lambda^{2}\right)\tilde{\mathbf{T}}^{0}
\end{eqnarray}
where $k_{0}^{\pm}=\pm\epsilon_{\pm}$. We have changed variable in
the third line: $\mathbf{k}\rightarrow\mathbf{k}+\boldsymbol{\lambda}_{s}$.
Above all, if $\mu_{\pm}=\mu\pm\mu_{5}$ and $\lambda_{\pm\mu}=\pm\lambda_{\mu}$,
then this current is
\begin{eqnarray}
\mathbf{j}=\frac{\Lambda\lambda_{i}}{2\pi^{2}}\tilde{\mathbf{T}}^{i}+\frac{\mu\lambda_{i}}{2\pi^{2}}\tilde{\mathbf{T}}^{i}+\frac{\mu\mu_{5}}{2\pi^{2}}\tilde{\mathbf{T}}^{0},
\end{eqnarray}
where the first term is the torsional chiral magnetic effect proposed in Ref. \cite{sumiyoshi2016prl}.

Let us turn to currents from the orbital moment. For $a=1,\thinspace2,\thinspace3$,
this current in the zero-temperature limit is
\begin{eqnarray}
\mathbf{j}_{s}^{\prime\left(1\right)} & = & s\int\frac{d^{3}k}{\left(2\pi\right)^{3}}f^{+}\partial_{\mathbf{k}}\frac{\left(k_{i}\mathbf{\tilde{T}}^{i}\cdot\widehat{\mathbf{k}-\mathbf{\boldsymbol{\lambda}}_{s}}\right)}{2\left|\mathbf{k}-\boldsymbol{\lambda}_{s}\right|}\nonumber \\
 & = & \frac{s\left(\mu_{s}-\lambda_{s0}\right)}{12\pi^{2}}\left(\lambda_{si}\mathbf{\tilde{T}}^{i}\right),
\end{eqnarray}
where contribution from negative-energy band is zero and we have performed
a partial integral and a variable change $\mathbf{k}\rightarrow\mathbf{k}+\boldsymbol{\lambda}_{s}$
in the second line.

For $a=0$, one can find
\begin{eqnarray}
\mathbf{j}_{s}^{\prime\left(2\right)} & = & s\int\frac{d^{3}k}{\left(2\pi\right)^{3}}f^{+}\partial_{\mathbf{k}}\frac{\left(k_{0}^{+}\mathbf{\tilde{T}}^{0}\cdot\widehat{\mathbf{k}-\mathbf{\boldsymbol{\lambda}}_{s}}\right)}{2\left|\mathbf{k}-\boldsymbol{\lambda}_{s}\right|}\nonumber \\
 & = & \frac{s\mu_{s}\left(\mu_{s}-\lambda_{s0}\right)}{12\pi^{2}}\left(\mathbf{\tilde{T}}^{0}\right).
\end{eqnarray}
Similarly, if $\mu_{\pm}=\mu\pm\mu_{5}$ and $\lambda_{\pm\mu}=\pm\lambda_{\mu}$,
this current becomes
\begin{equation}
\mathbf{j}_{s}^{\prime}=\frac{\mu}{6\pi^{2}}\left(\lambda_{i}\mathbf{\tilde{T}}^{i}\right)+\frac{\mu\mu_{5}}{3\pi^{2}}\mathbf{\tilde{T}}^{0}-\frac{\mu\lambda_{0}}{6\pi^{2}}\mathbf{\tilde{T}}^{0}.
\end{equation}

%%%%%%%%%%%%%%%%%%%%%%%%%%%%%%%%%%%%%%%%%%%%%%%%%%%%%%%%%%%%
%
\section{Liouville anomaly}\label{LiouA}

The Liouville equation says that the phase-space current is conserved.
However, the Berry curvature is singular at Weyl nodes and thus breaks
the Liouville equation, which is called Liouville anomaly. In this
section, we derive the Liouville equation in the language of differential
form \citep{dwivedi2013iop}. In addition, for clarity, we here use
the four-vector notation, with metric $\text{diag}\left(1,\thinspace-1,\thinspace-1,\thinspace-1\right)$
and neglect both $\xi_{\text{viel}}$ and $\xi_{\text{em}}$.

We define following one form
\begin{eqnarray}
-\eta_{H}=k_{i}dq^{i}+\left(\left|\mathbf{k}-\boldsymbol{\lambda}_{s}\right|+\lambda_{s0}\right)dt-w_{\mu}^{a}k_{a}dq^{\mu}-a_{sk}^{\mu}dk_{\mu}+A_{\mu}dq^{\mu},
\end{eqnarray}
and two-form $\omega_{H}=d\eta_{H}$, i.e.
\begin{eqnarray}
-\omega_{H} & = & \delta_{j}^{i}dk_{i}\wedge dq^{j}-\widehat{k-\lambda}_{s}^{i}dk_{i}\wedge dt\nonumber \\
 &  & -\left(k_{a}T^{a}+w_{\mu}^{i}dk_{i}\wedge dq^{\mu}-w_{\mu}^{0}\widehat{k-\lambda_{s}}^{i}dk_{i}\wedge dq^{\mu}\right)-\Omega_{s}+F,
\end{eqnarray}
where $s=\pm1$ for the chirality of Weyl fermions and $k_{0}=\left|\mathbf{k}-\boldsymbol{\lambda}_{s}\right|+\lambda_{s0}$.
$\Omega_{s}=\frac{1}{2}\Omega_{s}^{ij}dk_{i}\wedge dk_{j}$, $F$
and $T^{a}$ is the Berry curvature, electromagnetic tensor and torsion,
respectively. Because that $\omega_{H}=d\eta_{H}$, one would naively
expect that $d\omega_{H}=0$. However, this is not true. To appreciate
this point, let us calculate $d\omega_{H}$,
\begin{eqnarray}
-d\omega_{H} & = & -dT^{a}k_{a}-\partial_{\nu}w_{\mu}^{i}dq^{\nu}\wedge dk_{i}\wedge dq^{\mu}+\partial_{\nu}w_{\mu}^{0}\widehat{k-\lambda_{s}}^{i}dq^{\nu}\wedge dk_{i}\wedge dq^{\mu}\nonumber \\
 &  & -\left(-\widehat{k-\lambda_{s}}^{n}T^{0}dk_{n}+T^{i}dk_{i}\right)-d\Omega_{s}+dF\nonumber \\
 & = & -\left(dT^{a}k_{a}+d\Omega_{s}-dF\right),
\end{eqnarray}
where $dF=dT^{a}=0$. $d\Omega$ is not necessarily zero, but relates
to monopole charges. We then define a vector $V=\dot{q}^{i}\frac{\partial}{\partial q^{i}}+\dot{k}_{i}\frac{\partial}{\partial k_{i}}+\frac{\partial}{\partial t}$,
which is about translation along time: for an arbitrary function $g\left(q,\thinspace k,\thinspace t\right)$,
$Vg=\frac{d}{dt}g$.

Then, the Liouville equation is
\begin{eqnarray}
\mathcal{L}_{V}\Omega_{L} & = & \frac{1}{2!}d\Omega_{s}\wedge\left[\left(dk_{i}\wedge dq^{i}-\hat{k}^{i}dk_{i}\wedge dt-w_{\mu}^{i}dk_{i}\wedge dq^{\mu}+w_{\mu}^{0}\hat{k}^{i}dk_{i}\wedge dq^{\mu}\right)-k_{a}T^{a}-\Omega+F\right]^{2}\nonumber \\
 & = & \frac{1}{2!}d\Omega\left(k_{a}T^{a}-F\right)\wedge\left(k_{b}T^{b}-F\right),
\end{eqnarray}
where $\Omega_{L}=\frac{1}{3!}\omega_{H}^{3}\wedge dt$ is the phase-space
volume form and $\mathcal{L}_{V}$ is the Lie derivative of vector
$V$. $\mathcal{L}_{V}\Omega_{L}$ is now a top form, so, for convenience,
we employ the Hodge star operator to transform it to a scalar function,
i.e.
\begin{eqnarray}
\star\mathcal{L}_{V}\Omega_{L} & = & \star\left[\left(\frac{1}{2!}\right)^{4}\left(\frac{\partial}{\partial k_{l}}\Omega_{s}^{ij}dk_{l}\wedge dk_{i}\wedge dk_{j}\right)\left(k_{a}T_{\mu\nu}^{a}-F_{\mu\nu}\right)\left(k_{a}T_{\rho\sigma}^{a}-F_{\rho\sigma}\right)dq^{\mu}\wedge dq^{\nu}\wedge dq^{\rho}\wedge dq^{\sigma}\right]\nonumber \\
 & = & \left[\left(\frac{1}{2!}\right)^{4}\epsilon_{lij}\left(\frac{\partial}{\partial k_{l}}\Omega_{s}^{ij}\right)\epsilon^{\mu\nu\rho\sigma}\left(k_{a}T_{\mu\nu}^{a}-F_{\mu\nu}\right)\left(k_{a}T_{\rho\sigma}^{a}-F_{\rho\sigma}\right)\right]\nonumber \\
 & = & \frac{1}{8}\left(\partial_{\mathbf{k}}\mathbf{\tilde{\Omega}}_{s}\right)\epsilon^{\mu\nu\rho\sigma}\left(k_{a}T_{\mu\nu}^{a}-F_{\mu\nu}\right)\left(k_{b}T_{\rho\sigma}^{b}-F_{\rho\sigma}\right),
\end{eqnarray}
with $\left(\partial_{\mathbf{k}}\mathbf{\tilde{\Omega}}_{s}\right)=-2\pi s\delta\left(\mathbf{k}-\boldsymbol{\lambda}_{s}\right)$.
Thus, the Liouville equation is
\begin{equation}
\frac{\partial D}{\partial t}+\frac{\partial D\dot{\mathbf{q}}}{\partial\mathbf{q}}+\frac{\partial D\dot{\mathbf{k}}}{\partial\mathbf{k}}=-\frac{\pi s\epsilon^{\mu\nu\rho\sigma}}{4}\delta^{3}\left(\mathbf{k}-\boldsymbol{\lambda}_{s}\right)\left(k_{a}T_{\mu\nu}^{a}-F_{\mu\nu}\right)\left(k_{b}T_{\rho\sigma}^{b}-F_{\rho\sigma}\right).
\end{equation}
Because of $k_{0}=\left|\mathbf{k}-\boldsymbol{\lambda}_{s}\right|+\lambda_{s0}$,
this delta function implies $k_{0}=\lambda_{s0}$. In addition, by
inserting distribution function back, the equation above becomes
\begin{equation}
\partial_{\mu}j_{s}^{\mu}=-s\frac{\epsilon^{\mu\nu\rho\sigma}}{32\pi^{2}}\left(\lambda_{sa}T_{\mu\nu}^{a}-F_{\mu\nu}\right)\left(\lambda_{sb}T_{\rho\sigma}^{b}-F_{\rho\sigma}\right).
\end{equation}

%%%%%%%%%%%%%%%%%%%%%%%%%%%%%%
%\section{ derivation of Eq. (11) from the chiral Landau levels}\label{D}

%Let us turn to consider Weyl semimetals with screw dislocations along the $z$-axis. The Weyl nodes with chirality $s$ locate at $s\lambda_{\mu}$ with $s=\pm1$. For simplicity, we set $\lambda_{\mu}=(0,\,0,\,0,\,-\lambda_{z})$ and $\lambda_{z}>0$.
%The displacement vector is assumed to be along the $z$-axis, so only one component of the vielbeins survives $e^{3}_{\mu}=\frac{1}{2}(0,\,-\tilde{T}y,\,\tilde{T}x,\,0)$.
%Note that the surface density of the Burgers vector fields rather than itself is constant. Consequently, the zeroth Landau levels near $p_{z}=s\lambda_{z}$ are $p_z-s\lambda_{z}$. Turning on an electric field along the $z$-axis, charges are pumped up from the Dirac sea and extra particles are "produced". Specifically, the level degeneracy is roughly $\frac{\lambda_{z}\tilde{T}}{2\pi}$ in the vicinity of $p_{z}=s\lambda_z$ and the variation of momentum is, $\triangle p_z=E\triangle t$. Hence, the total variation of charge density is given as $\partial_{t} j^0=\frac{1}{2\pi^2}\lambda_{z}\tilde{T}E$, whose covariant form is Eq.~(11) in the main text.

%%%%%%%%%%%%%%%%%%%%%%%%%%%%%%%%%%%%%%%%%%%%%%%%%%%%%%%%%%%%
\section{``Landau Levels'' induced by screw dislocations}\label{LLs}

In this section, we derive the ``Landau levels'' for Weyl fermions
under screw dislocations. From the action in Eq. $\left(1\right)$
in the main text, one obtains the corresponding equation of motion
\begin{equation}
\frac{1}{2}\left[\left(\det e_{\beta}^{b}\right)\gamma^{a}e_{a}^{\mu}\left(i\partial_{\mu}\Psi\right)+\gamma^{a}i\partial_{\mu}\left(\det e_{\beta}^{b}e_{a}^{\mu}\Psi\right)\right]-\left(\det e_{\beta}^{b}\right)\lambda_{a}\gamma^{a}\gamma_{5}\Psi=0.
\end{equation}
For simplicity, we consider screw dislocations with the displacement
vector along $z$-axis
\begin{equation}
e_{\mu}^{a}=\delta_{\mu}^{a}+w_{\mu}^{a},
\end{equation}
with $w_{\mu}^{3}=\frac{1}{2}\left(0,\thinspace-\tilde{T}y,\thinspace\tilde{T}x,\thinspace0\right)$
and $w_{\mu}^{a}=0$ for $a\neq3$. That is, the torsional magnetic
field $\tilde{T}$ is along the $z$-axis.

The Hamiltonians for the right-handed fermions ($H_{R}$) and the
left-handed fermions ($H_{L}$) are thus given as
\begin{equation}
H_{R}=\left(\begin{array}{cc}
\left(p_{z}-\lambda_{z}\right) & \left(p_{x}+\frac{1}{2}\tilde{T}yp_{z}\right)-i\left(p_{y}-\frac{1}{2}\tilde{T}xp_{z}\right)\\
\left(p_{x}+\frac{1}{2}\tilde{T}yp_{z}\right)+i\left(p_{y}-\frac{1}{2}\tilde{T}xp_{z}\right) & -\left(p_{z}-\lambda_{z}\right)
\end{array}\right),
\end{equation}
and
\begin{equation}
H_{L}=-\left(\begin{array}{cc}
\left(p_{z}\boldsymbol{+}\lambda_{z}\right) & \left(p_{x}+\frac{1}{2}\tilde{T}yp_{z}\right)-i\left(p_{y}-\frac{1}{2}\tilde{T}xp_{z}\right)\\
\left(p_{x}+\frac{1}{2}\tilde{T}yp_{z}\right)+i\left(p_{y}-\frac{1}{2}\tilde{T}xp_{z}\right) & -\left(p_{z}\boldsymbol{+}\lambda_{z}\right)
\end{array}\right).
\end{equation}
It implies that Weyl fermions under a momentum-dependent magnetic
field $\tilde{T}p_{z}$, which entirely differ from the cases of magnetic
fields and axial magnetic fields.

Because the Hamiltonian commutes with $\hat{p}_{z}$, the quantum
number $p_{z}$ can be used to label eigenstates and the Hamiltonian
can be recast as
\begin{equation}
H_{R}=\begin{cases}
\begin{array}{c}
\left(\begin{array}{cc}
\left(p_{z}-\lambda_{z}\right) & \sqrt{\left|2\tilde{T}p_{z}\right|}\hat{A}^{\dagger}\\
\sqrt{\left|2\tilde{T}p_{z}\right|}\hat{A} & -\left(p_{z}-\lambda_{z}\right)
\end{array}\right)\\
\left(\begin{array}{cc}
\left(p_{z}-\lambda_{z}\right) & \sqrt{\left|2\tilde{T}p_{z}\right|}\hat{A}\\
\sqrt{\left|2\tilde{T}p_{z}\right|}\hat{A}^{\dagger} & -\left(p_{z}-\lambda_{z}\right)
\end{array}\right)
\end{array} & \begin{array}{c}
\text{for}\thinspace\thinspace\text{ }\tilde{T}p_{z}>0\\
\text{for}\thinspace\thinspace\text{ }\tilde{T}p_{z}<0
\end{array}\end{cases},
\end{equation}
where $\hat{b}_{x}=-i\partial_{x}-i\frac{1}{2}\tilde{T}xp_{z}$, $\hat{b}_{y}=-i\partial_{y}-i\frac{1}{2}\tilde{T}yp_{z}$,
\begin{equation}
\hat{a}_{x\left(y\right)}=\begin{cases}
\begin{array}{c}
\frac{1}{\sqrt{\left|\tilde{T}p_{z}\right|}}\hat{b}_{x\left(y\right)}\\
\frac{1}{\sqrt{\left|\tilde{T}p_{z}\right|}}\hat{b}_{x\left(y\right)}^{\dagger}
\end{array} & \begin{array}{c}
\text{for}\thinspace\thinspace\text{ }\tilde{T}p_{z}>0\\
\text{for}\thinspace\thinspace\text{ }\tilde{T}p_{z}<0
\end{array},\end{cases}
\end{equation}
and
\begin{equation}
\hat{A}=\frac{\hat{a}_{x}+i\hat{a}_{y}}{\sqrt{2}}.
\end{equation}
It is straightforward to verify that $\left[\hat{A},\thinspace\hat{A}^{\dagger}\right]=1$
and $\left[\hat{A},\thinspace\hat{A}\right]=0$. The square of Hamiltonian is
\begin{equation}
H_{R}^{2}|_{\tilde{T}p_{z}>0}=\left(\begin{array}{cc}
\left(p_{z}-\lambda_{z}\right)^{2}+\left|2\tilde{T}p_{z}\right|\hat{A}^{\dagger}\hat{A} & 0\\
0 & \left|2\tilde{T}p_{z}\right|\left(\hat{A}^{\dagger}\hat{A}+1\right)+\left(p_{z}-\lambda_{z}\right)^{2}
\end{array}\right)
\end{equation}
and
\begin{equation}
H_{R}^{2}|_{\tilde{T}p_{z}<0}=\left(\begin{array}{cc}
\left(p_{z}-\lambda_{z}\right)^{2}+\left|2\tilde{T}p_{z}\right|\left(\hat{A}^{\dagger}\hat{A}+1\right) & 0\\
0 & \left|2\tilde{T}p_{z}\right|\hat{A}^{\dagger}\hat{A}+\left(p_{z}-\lambda_{z}\right)^{2}
\end{array}\right).
\end{equation}
Thus, the dispersion relation for the right-handed fermions can be calculated
\begin{equation}
\mathcal{E}_{R}=\begin{cases}
\begin{array}{c}
\begin{cases}
\begin{array}{c}
p_{z}-\lambda_{z}\\
-\left(p_{z}-\lambda_{z}\right)
\end{array} & \begin{array}{c}
\text{for}\thinspace\thinspace\text{ }\tilde{T}p_{z}>0\\
\text{for}\thinspace\thinspace\text{ }\tilde{T}p_{z}<0
\end{array}\end{cases}\\
\pm\sqrt{\left(p_{z}-\lambda_{z}\right)^{2}+2\left|n\tilde{T}p_{z}\right|}
\end{array} & \begin{array}{c}
n=0\\
n^{2}\geq1
\end{array},\end{cases}
\end{equation}
where $n$ is integer. Similarly, for the left-handed fermions, one has
\begin{equation}
\mathcal{E}_{L}=\begin{cases}
\begin{array}{c}
\begin{cases}
\begin{array}{c}
-\left(p_{z}+\lambda_{z}\right)\\
\left(p_{z}+\lambda_{z}\right)
\end{array} & \begin{array}{c}
\text{for}\thinspace\thinspace\text{ }\tilde{T}p_{z}>0\\
\text{for}\thinspace\thinspace\text{ }\tilde{T}p_{z}<0
\end{array}\end{cases}\\
\pm\sqrt{\left(p_{z}+\lambda_{z}\right)^{2}+2\left|n\tilde{T}p_{z}\right|}
\end{array} & \begin{array}{c}
n=0\\
n^{2}\geq1
\end{array}.\end{cases}
\end{equation}
One clearly finds a chiral zeroth Landau level for the left- or right-handed Weyl fermions, as shown in Fig.~\ref{fig1} above.
\begin{figure}
\includegraphics[scale=0.6]{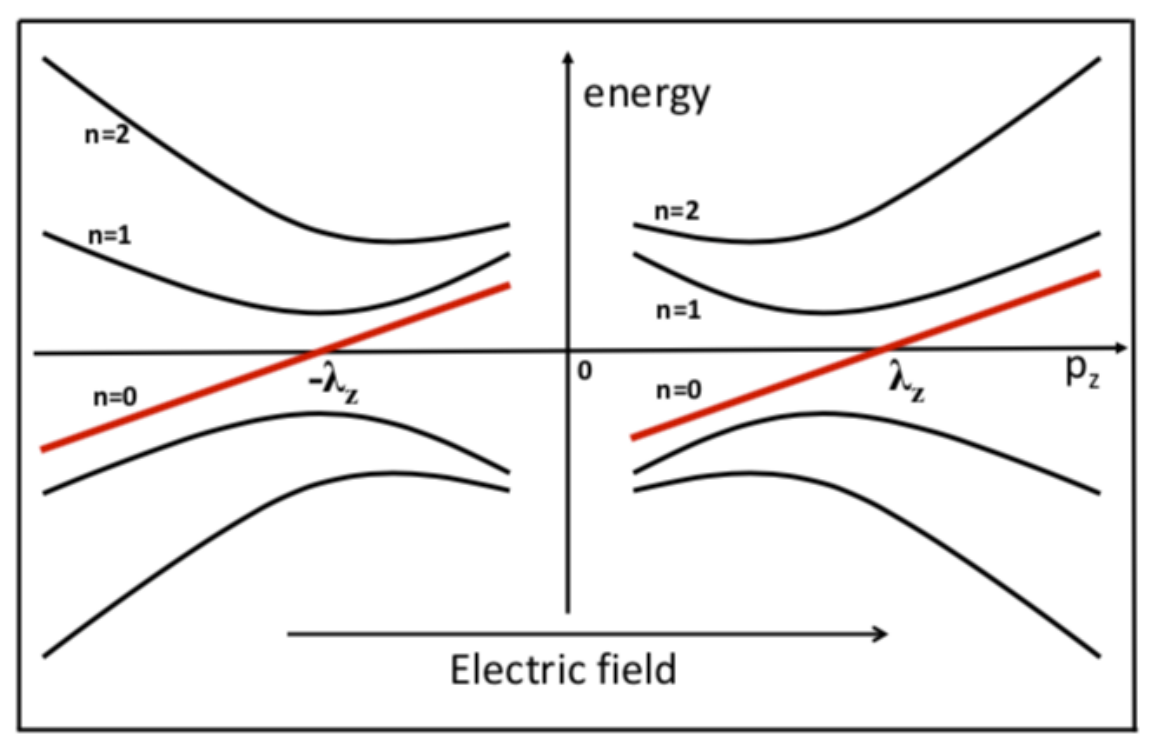}\caption{Schematic picture of the Landau levels due to screw dislocations. The red lines refer to the zeroth chiral Landau level for the right-handed Weyl fermions near $p_{z}=\lambda_{z}$ and the left-handed Weyl fermions near $p_{z}=-\lambda_{z}$, respectively\label{fig1}. }
\end{figure}
%
%%%%%%%%%%%%%%%%%%%%%%%%%%%%%%%%%%%%%%%%%%%%%%%%%%%%%%%%%%%%
%
\section{Modified Gordon's identity~\label{MGI}}

The well-known Gordon's decomposition is valid for massive Dirac fermions.
Therefore, in this section, we derived a modified Gordon's identity, which also holds for massless Weyl fermions.

Assume $|u_{+}\left(p\right)\rangle$ is the eigenfunction of Weyl's equation,
\begin{eqnarray}
\mathbf{p}\cdot\mathbf{\sigma}|u_{+}\left(p\right)\rangle=+\left|\mathbf{p}\right||u_{+}\left(p\right)\rangle.
\end{eqnarray}
From the identity for Pauli's matrices
\begin{eqnarray}
\left[\sigma^{i},\thinspace\sigma^{j}\right]=2\sigma^{i}\sigma^{j}-2\delta^{ij}=-2\sigma^{j}\sigma^{i}+2\delta^{ij},
\end{eqnarray}
one can thus obtain
\begin{align*}
 & \frac{1}{2}\langle u_{+}\left(p^{\prime}\right)|\left[\sigma^{i},\thinspace\sigma^{j}\right]\left(p^{\prime}-p\right)_{j}|u_{+}\left(p\right)\rangle\\
= & \langle u_{+}\left(p^{\prime}\right)|\left[\left(-\sigma^{j}\sigma^{i}+\delta^{ij}\right)p_{j}^{\prime}-\left(\sigma^{i}\sigma^{j}-\delta^{ij}\right)p_{j}\right]|u_{+}\left(p\right)\rangle\\
= & \langle u_{+}\left(p^{\prime}\right)|\left[-\sigma^{j}\sigma^{i}p_{j}^{\prime}-\sigma^{i}\sigma^{j}p_{j}+\delta^{ij}\left(p_{j}^{\prime}+p_{j}\right)\right]|u_{+}\left(p\right)\rangle\\
= & \langle u_{+}\left(p^{\prime}\right)|\left(\left|\mathbf{p}^{\prime}\right|+\left|\mathbf{p}\right|\right)\sigma^{i}|u_{+}\left(p\right)\rangle-\left(\mathbf{p}^{\prime}+\mathbf{p}\right)^{i}\langle u_{+}\left(p^{\prime}\right)|u_{+}\left(p\right)\rangle,
\end{align*}
with $p_{\mu}=\left(p_{0},\thinspace-\mathbf{p}\right)$. Thus the
modified Gordon's identity for Weyl fermions is of form
\begin{eqnarray}
\langle u_{+}\left(p^{\prime}\right)|\sigma^{i}|u_{+}\left(p\right)\rangle & = & \frac{1}{\left(\left|\mathbf{p}^{\prime}\right|+\left|\mathbf{p}\right|\right)}[-i\epsilon^{ijk}\langle u_{+}\left(p^{\prime}\right)|\left(\mathbf{p}^{\prime}-\mathbf{p}\right)^{j}\sigma^{k}|u_{+}\left(p\right)\rangle+\left(\mathbf{p}^{\prime}+\mathbf{p}\right)^{i}\langle u_{+}\left(p^{\prime}\right)|u_{+}\left(p\right)\rangle].
\end{eqnarray}
That is, for the right-handed Weyl fermions, we have found
\begin{eqnarray}
\langle u_{+}\left(p\right)|\sigma^{i}|u_{+}\left(p-\triangle p\right)\rangle & = & -i\epsilon^{ijk}\left[\langle u_{+}\left(p\right)|\sigma^{k}|u_{+}\left(p-\triangle p\right)\rangle\right]\frac{\left(\triangle\mathbf{p}\right)^{j}}{\left(\left|\mathbf{p}\right|+\left|\mathbf{p}-\triangle\mathbf{p}\right|\right)}\nonumber \\
 &  & +\langle u_{+}\left(p\right)|\frac{\left(2\mathbf{p}-\triangle\mathbf{p}\right)^{i}}{\left(\left|\mathbf{p}\right|+\left|\mathbf{p}-\triangle\mathbf{p}\right|\right)}|u_{+}\left(p-\triangle p\right)\rangle,
\end{eqnarray}
which leads to following equation by iterating
\begin{align}
 & \langle u_{+}\left(p\right)|\sigma^{i}|u_{+}\left(p-\triangle p\right)\rangle\nonumber \\
= & \langle u_{+}\left(p\right)|u_{+}\left(p-\triangle p\right)\rangle[\frac{\left(2\mathbf{p}-\triangle\mathbf{p}\right)^{i}}{\left(\left|\mathbf{p}\right|+\left|\mathbf{p}-\triangle\mathbf{p}\right|\right)}-i\epsilon^{ijk}\frac{\left(2\mathbf{p}-\triangle\mathbf{p}\right)^{k}}{\left(\left|\mathbf{p}\right|+\left|\mathbf{p}-\triangle\mathbf{p}\right|\right)}\frac{\left(\triangle\mathbf{p}\right)^{j}}{\left(\left|\mathbf{p}\right|+\left|\mathbf{p}-\triangle\mathbf{p}\right|\right)}]+\mathcal{O}\left(\triangle p^{2}\right)
\end{align}
Because of the expansion $\left|\mathbf{p}-\triangle\mathbf{p}\right|=\left|\mathbf{p}\right|-\hat{\mathbf{p}}\cdot\triangle\mathbf{p}+\mathcal{O}\left(\triangle\mathbf{p}^{2}\right),$
one can find
\begin{eqnarray}
\frac{\left(2\mathbf{p}-\triangle\mathbf{p}\right)^{i}}{\left(\left|\mathbf{p}\right|+\left|\mathbf{p}-\triangle\mathbf{p}\right|\right)} & = & \frac{\hat{\mathbf{p}}^{i}+\hat{\mathbf{p}}^{\prime i}}{2}+\mathcal{O}\left(\triangle p^{2}\right)
\end{eqnarray}
where $\mathbf{p}^{\prime}=\mathbf{p}-\triangle\mathbf{p}.$ That is, equation above becomes
\begin{align}
\langle u_{+}\left(p\right)|\sigma^{i}|u_{+}\left(p-\triangle p\right)\rangle= & \langle u_{+}\left(p\right)|u_{+}\left(p-\triangle p\right)\rangle\left[\frac{\hat{\mathbf{p}}^{i}+\hat{\mathbf{p}}^{\prime i}}{2}-i\epsilon^{ijk}\frac{\left(\hat{\mathbf{p}}^{k}+\hat{\mathbf{p}}^{\prime k}\right)}{2}\frac{\left(\triangle\mathbf{p}\right)^{j}}{2\left|\mathbf{p}\right|}\right]+\mathcal{O}\left(\triangle p^{2}\right).
\end{align}
\end{widetext}
\end{appendix}

%%%%%%%%%%%%%%%%%%%%%%%%%%%%%%%%%%%%%%%%%%%%%%%%%%%%
%%%%%%%%%%%%%%%%%%%%%%%%%%%%%%%%%%%%%%%%%%%%%%%%%%%%
\bibliographystyle{apsrev4-1}
%\bibliography{AMWF}

\begin{thebibliography}{72}
\bibitem{weakdisp} In this work, we mainly focus on the weak displacement field regime ( small $u^{a}$ compared to the crystal constants) away from the dislocation cores, in which the perturbative approach is sufficient to investigate the responses from torsions.
\end{thebibliography}

\end{document}